# Loading Arbitrary Knowledge Bases in Matrix Browser

Saqib Saeed[1] and Christoph Kunz[2]
[1]Bahria University Islamabad, Pakistan.
[2]Fraunhofer Institute for Industrial Engineering Stuttgart Germany.
*Saqib@bci.edu.pk, christoph.kunz@iao.fhg.de*

## Abstract

*This paper describes the work done on Matrix Browser, which is a recently developed graphical user interface to explore and navigate complex networked information spaces. This approach presents a new way of navigating information nets in "windows explorer" like widget. The problem on hand was how to export arbitrary knowledge bases in Matrix Browser. This was achieved by identifying the relationships present in knowledge bases and then by forming the hierarchies from this data and these hierarchies are being exported to matrix browser. This paper gives solution to this problem and informs about implementation work.*

## 1. Introduction

Matrix Browser is a graphical user interface that uses adjacency matrix approach to represent information in form of graph structures. In adjacency matrix approach, nodes are defined at horizontal and vertical axis. The relationships between different nodes are determined by corresponding cell where respective row and column intersects each other. The direction of the relationship is defined by arrow like shape on the square box representing the relationship. In order to give user an option to explore between hierarchies system has to derivate new associations based on the associations present in the ancestor nodes. Some times cell does not show real link present between different nodes but it shows a symbol, which tells that by exploring this you can reach to the real link present between the nodes that is present at lower level of the hierarchy.

The hierarchies present on the axis can be explored directly by expanding and collapsing interactive trees. This expanding and collapsing of trees takes place on the both axis of matrix browser. This type of information navigation is easy to realize. The relationship between the two selected nodes should always be consistent. In order to achieve it, there are two important points to ponder. First of all the nodes which are invisible due to the collapsed state of their upper nodes and secondly the transitive relationships on the basis of which the relationship between two nodes can be propagated upwards or downwards in the hierarchy. There are different types of relationships present in the cells of the matrix. If a cell in the matrix browser is marked then it would have one of the following types of relationships.

### 1.1 Explicit relationship

It means that the relationship between the nodes is explicitly defined in the underlying graph.

### 1.2 Implicit relationship

It means that the relationship has been derived from the underlying graph with the help of some reference mechanism.

### 1.3. Hidden relationship

It means that the explicit relationship between nodes is present in the hierarchy at some lower level but currently is not visible due to the collapsed state of upper nodes.

Different symbols are used in order to represent different types of relationships in the cells of the matrix browser. The hidden and implicit relationships can be expanded like the nodes in a tree widget by clicking them. The symbol () represents the existence of a relationship between the nodes that are collapsed. This relationship can be explicit or implicit. When this symbol is clicked the hierarchies expand.

The symbol () shows that the nodes are in expanded form and clicking on this symbol can minimize the hierarchies. When clicking this symbol it will turn to this shape ( ), which means that the hierarchy can be expanded, and there exists an implicit relationship somewhere down in the hierarchy.

The symbol ( ) denotes that there is an explicit relationship present between these nodes. The graph structure at the top left corner in the matrix browser denotes the relationships of a selected node. The selected node is represented in centre whereas all the nodes to which it has any relationship form a net around it.

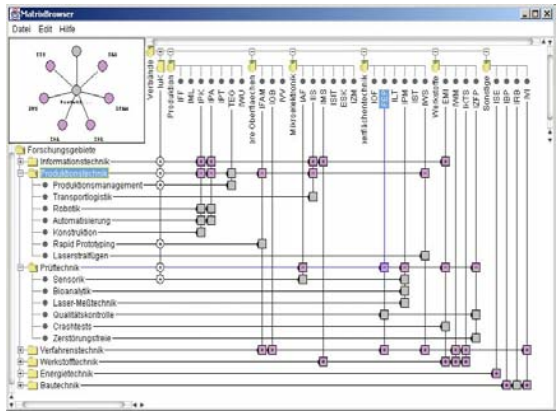

**Figure 1. Matrix Browser Snapshot**

In this way the information can easily be explored until the bottom and one can go from higher-level concepts to lower level nodes. [1]. The Figure 1 gives an overview of the representation in the matrix browser.

## 2. Problem Definition

A store of knowledge about a domain represented in machine process able form, which may be rules (in which case the knowledge base may be considered a rule base), facts, or other representations.

In order to enhance the functionality of matrix browser the import of arbitrary knowledge bases is planned. There are following two main problems in loading arbitrary knowledge bases into matrix browser.

- Extraction of Data
- Hierarchy Formation

The data present in knowledge base is in form of classes, objects, and different relationships between classes and objects. The relationships can be inherited or defined in objects, and classes. The nodes are interconnected with each other. The main area of concern here is that which type of nodes would come together in a tree and what are the relationships on the basis of which the nodes can be combined together in a tree.

Secondly after successfully forming the trees from underlying data how the hierarchies would be constructed from these trees.

## 3. Extraction of Data

Firstly it is detected that which nodes are root nodes of data defined in knowledge base so that it is clear that how many trees would form from the knowledge base. When these roots are identified then relationships between these nodes are extracted, every node holds its children.

The trees are build with "subclass of" relationships between classes which means that if two classes have subclass of relationships between them then they can be inserted in a tree where parent class would form the parent node of the tree and class which is subclass of that would be child of that node and type of relationship is also returned which in this case would be the 'subclass of' and type of relationship will be displayed as tool tip of node when it is displayed in Matrix Browser.

The other possibility of making tree is when an object is instantiated of a class then there exists an "instance of" relationship. The name of instance would be the child node and class to which it belongs would be the parent node and type of relationship would be 'instance of' and it is displayed as tool tip of the node when it is displayed in the matrix browser.

Another possibility exists for "Part of" relationships the object or classes that have "Part of" attribute would also form trees. The 'Part of' relationship can be single valued or multi valued in both cases this is extracted to form hierarchies. Here the object which has "Part of" property would form child node and value of that attribute would be parent node and here type would be 'Part of ' which would be displayed as tool tip of child node.

All the remaining relationships are analysed which can be either single valued relationships, multi valued relationships, meta relationships or inherited relationships. The single valued relationships are those relationships which have single value assigned to them while in case of multi valued attributes the attribute can have more than one values assigned to it. Whereas meta relations are relations which are given to classes to implement the conditions that result of attribute should belong to a specific class. All inherited relationships also needed to be extracted these are those relationships which are inherited by the class to which objects belong.

Among the inherited relations single valued, multi valued or the meta relations can exist.

## 4. Hierarchy Formation

When trees are formed then the following cases can appear while extracting the hierarchies from the underlying concept graph.

### 4.1 Case 1

If the trees are disjoint then there is no need to process them they can be extracted simply as it.

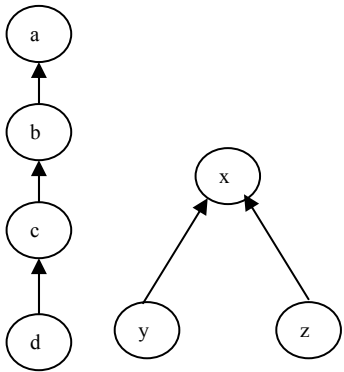

**Figure 2. Disjoint Trees**

This situation is described in Figure 2 that two trees are present in knowledge base and both of them are disjoint means that they do not have a node which is common to both trees and there is no direct relationship with each other then in this case these trees present in knowledge base are directly extracted into the matrix browser

### 4.2 Case 2

If a situation arises where trees overlap each other then the part, which is common in both trees, will come in both trees separately and will have identity relationship among them.

Suppose if there is a case present as in Figure 3. There are different hierarchies present in a tree extracted from knowledge base then it should be divided in three different hierarchies. In the Figure 4 it can be seen that node 'b' has two nodes as its parents, the node 'a' and node 'c' so tree is further divided into two trees. The one tree will have the node 'a' as root node whereas second tree will have node 'f' as the root node. The same is the case with node 'c'. It has also two parents in form of node 'f' and node 'a'. So tree mentioned in Figure 3 would have three hierarchies in it as shown in Figure 4.

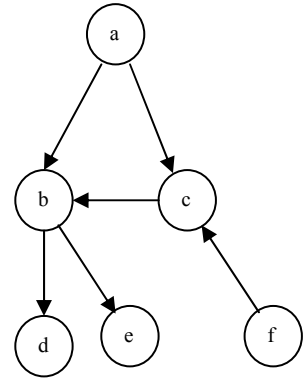

**Figure 3. Overlapped Tree**

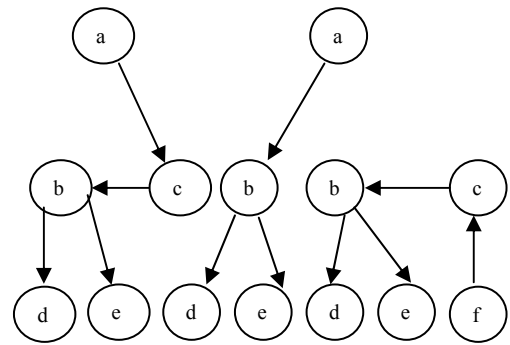

**Figure 4. Separation of Overlapped Tree**

### 4.3 Case 3

If a circular type of tree is extracted from the knowledge base then one can just copy last node additionally and making that node as a new root of hierarchy.

Suppose there are four nodes named a, b, c and d they have part of relationship between them as the situation described in Figure 5 d is "part of" c and c is "part of" b and b is "part of" a and in the end a is "part of" d. In order to convert them into a hierarchy last node will be copied additionally and it will become root of new hierarchy and the same copy of this node will also be present as last node in the hierarchy. [2]

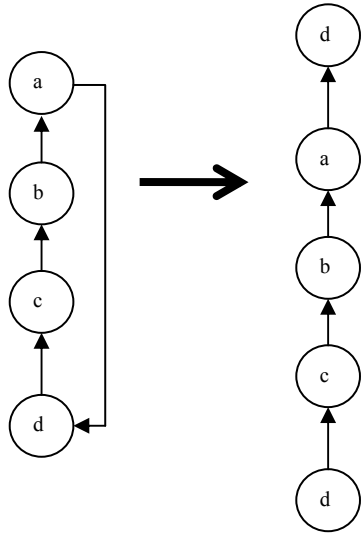

**Figure 5. Circular Tree**

## 5. Implementation

The FLORA-2 implementation of F-Logic is selected. This runs under the XSB engine. This implementation is selected because it is open source. Secondly it is more stable implementation available at the moment. The other choices were "Tripple" and "Onto Edit". The "Tripple" is not able to do complex processing whereas "Onto Edit" is not available as open source and the FLORA-2 syntax is only limited to this editor.[3].

When the Matrix Browser loads an arbitrary Flora-2 knowledge base there is set of queries defined which is embedded in the arbitrary knowledge base. This set of queries is responsible for extracting all the data from knowledgebase and the transferring this data to Java environment by using Interprolog API. These trees are then processed in Java environment to form hierarchies. Now all relationships are analyzed except special relationships (partof, subclassf, instanceOf). If value for a relationship is present in hierarchy at the other axis, then it is represented in the form of a cell at the intersection of both nodes, and relationship's name is represented as tool-tip and if value of the relationship is not present as a node in network then this relationship becomes attribute of that node and represented as a tool tip of source node (from where the relationship originates).

## 6. Summary and Future Work

After the completion of Proposed work Matrix Browser is capable of loading an arbitrary Flora-2 knowledge base and representing generated hierarchies in Matrix Browser. In order to further improve the functionality of Matrix Browser, task of importing DAML+OIL ontologies can be implemented. As a result the Matrix Browser shall be capable of importing DAML+OIL concept graphs, which is a widely used language for describing ontological data.

The import of XML topic maps can also be made possible. The DAML+OIL and XML topic maps are two standards for describing the ontological information. Both standards are used by different group of people. So in this way both standards can be handled. [4]

At present the Matrix Browser is only capable of just querying knowledge base once. This work can be enhanced by implementing a framework by which user can define semantic queries from inside the visualization and sending the queries to knowledge base for information retrieval, then retrieved data is displayed in the matrix browser.

The matrix browser can be developed into a web service. The matrix browser can work on client side. The inference engine can be encapsulated in J2EE application server. The inference engine and the DAML+OIL file can serve as two tiers at server.